\documentclass{article}
\usepackage{graphicx}
\usepackage{booktabs}
\usepackage{epstopdf}

\begin{document}

\title{ 
DESIGN OF  PRE-DUMPING RING SPIN ROTATOR WITH A POSSIBILITY OF HELICITY SWITCHING FOR POLARIZED POSITRON  AT THE ILC \thanks{ Work supported by  the German Federal Ministry of education and research, Joint  Research project  R$\&$D  Accelerator "Spin Management", contract N 05H10CUE}}

\author{L.I. Malysheva\thanks{larisa.malysheva@desy.de
, L.I.Malysheva@salford.ac.uk}, O.S. Adeyemi,V. Kovalenko, G.A. Moortgat-Pick,\\
A. Ushakov, Hamburg University,  Hamburg, Germany\\
S. Riemann, F. Staufenbiel,  DESY, Zeuthen, Germany\\
A. Hartin, B. List, N.J. Walker, DESY, Hamburg, Germany\\
}

\maketitle

\begin{abstract}
     The use of  polarized beams enhance the possibility of the precision measurements at the International Linear Collider (ILC) \cite{1}.
     In order  to preserve the  degree of polarization during beam transport  spin rotators  are included in the current TDR ILC lattice \cite{2}. In this report some advantages of using  a combined spin rotator/spin flipper section  are discussed. A few possible lattice designs of spin flipper developed  at DESY in 2012 are presented.
\end{abstract}

\section{Introduction}
The importance of beam polarization  for the ILC experiments  can be illustrated by fact that  the effective luminosity is increasing by  approximately $ 50\%$  in the case of both  beam  polarized \cite{1}.
   Furthermore a suitable combination of  polarized electron and positron  beams suppresses significantly unwanted background processes and enhances  signal rates.

 There are two important aspects which should be taken into account for polarized beams.
 The first one is a  delivery  of polarized beams from the source to the interaction point. The spin transport  for the different areas of  the ILC were already studied \cite{3,4} and  the installation  of spin rotators before and after Damping Ring was recommended.  The examples of possible layouts of spin-rotators for the ILC can be found in \cite{5,6}. The second problem arising from the presence of polarized beams is the requirement of fast helicity reversal. The helicity pattern of the electron beam can be adjusted by changing the helicity of the laser. For the positron beam this is a non-trivial  task, as the polarization of  the positron beam  depends on helicity of the undulator. The possibility of spin manipulation was considered at \cite{7} where two post-damping ring spin rotators were included.
  On the other side, the  spin manipulation of post damping ring beams is limited  by  the emittance preservation constraints. In addition the spin rotator used in the  TDR design cannot provide a fast helicity reversal in the time scale desirable for the ILC, i.e.  from train to train.

 The idea of using   a pre-damping ring  spin rotator section for some beam  helicity manipulations has been already suggested in \cite{8},  but no detailed lattice   was produced.  Meanwhile, the  layout of the Central Region of the ILC provides enough space before Damping Ring  for a combined spin rotation with  a possibility of  quick switch between two helicities.  A possible  layout of  a pre-damping ring spin rotator/spin flipper section  is presented  below.
\begin{figure}[htb]
   \centering
   \includegraphics*[width=65mm,height=38mm]{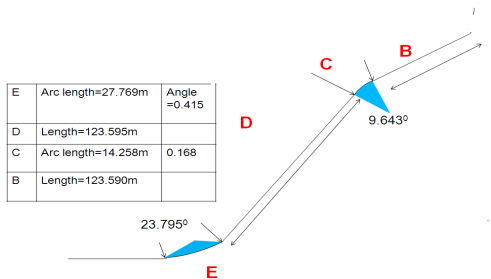}
   \caption{Schematic layout of  new PLTR section}
   \label{f1}
\end{figure}

\section{The ILC Pre-Damping Ring Spin Rotator}


The Positron  Linac To Damping Ring (PLTR) is a section of the ILC  transport  positron beam to the Damping Ring (DR).
The   schematic layout of  the PLTR is given in Fig.~\ref{f1}. It serves for the extraction of the positrons from the Positron Linac Booster, energy compression and spin rotation.

 In general, the  desirable spin rotation can be produced by spin precession around the field direction. In  the dipole field  the  rate of spin precession  is  directly proportional to the orbit deflection angle  $\theta_{orbit}$ while  in  the solenoidal  field spin precession rate is directly  proportional to the field $B_{z}$ and the length $L_{sol}$ of solenoid  and inversely proportional to the the magnetic rigidity $B \rho$. At 5 GeV the  orbital deflection angle of $7.929^0$  rotates  spins by $90^{0}$.  In section E the spin rotation   from  the longitudinal to the transversal direction is done by the means of horizontally bending dipoles with the total orbital rotation angle of $23.795^0= 3 \times 7.929^0$ which corresponds to $ \frac{3\pi }{2}$ of spin rotation.

The  total length of section D is 123.595 m. The  suggested combined spin flipper/spin rotator design  is  only 80  m long. A new  modified section D   can  fulfill two tasks simultaneously, namely  spin rotation and   train-by train helicity reversal. The energy compression  in section  D matches  the beam energy spread to the DR acceptance.  Then  the transversal  beam polarization  can be rotated to vertical  in the solenoid  with  a field integral of  26.18 [T m]. Two different superconducting solenoids design were considered: 8.32 m long solenoid  with an integrated field of 26.18 $[T \cdot m ]$ (Solenoid 1) and  a shorter 5m long superconducting solenoid with  integrated field of 26.2 $[T \cdot m ]$ (Solenoid 2). The pre-damping ring position of the spin- rotator makes the emittance preservation constrains less  challenging.
 \begin{figure*}[htb]
    \centering
    \includegraphics*[width=110mm,height=35mm]{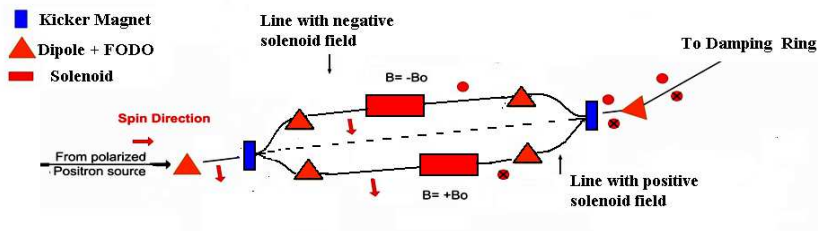}
    \caption{ The schematic layout of the positron transport to Damping Ring  with a  two parallel lines spin rotator section.}
    \label{f2}
\end{figure*}

 The suggested  combined spin flipper/spin rotator  consists of two parallel  beam lines for spin rotation equipped with  two solenoids of  opposite polarities, i.e. setting the spin parallel  (one beam line) or antiparallel (second beam line) to the field in the Damping Ring, Fig.~\ref{f2}.  This  spin- flipper design   is based on the concept of  branch splitter/merger used for the post-damping ring  positron lines \cite{9}. The first lattice cell is an irregular FODO cell which  include fast kickers and separate the branches horizontally.  The  total length of the  splitter section is approximately 26 m  in order to fit the  available space, 2m  of two  horizontal branches separation is  taken. The shortening of the splitter section is achieved by using stronger bending magnets.

 Each branch  consists of a first order achromat  FODO dogleg, a solenoid section and another dogleg to recombine the line back to the design orbit. The achromat design assures that no dispersion suppressors would be required.  The simple solenoid rotator design is considered, similar to the one used in \cite{7}.
The advantage of this  design is  the possibility of quick  and random switching between two helicities for the positrons. In order to save some transversal space an asymmetry can be introduced in  the relative  position of  solenoids at two branches.

\subsection{Symmetric Design}

\begin{figure}[htb]
   \centering
   \includegraphics*[width=75mm,height=34mm]{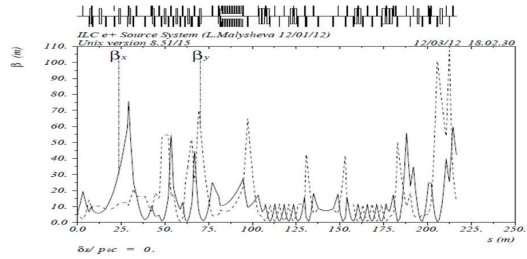}
   \caption{ Complete PLTR section including one of spin rotator branch  with solenoid 1.}
   \label{f3}
\end{figure}

\begin{figure}[htb]
   \centering
   \includegraphics*[width=80mm,height=36mm]{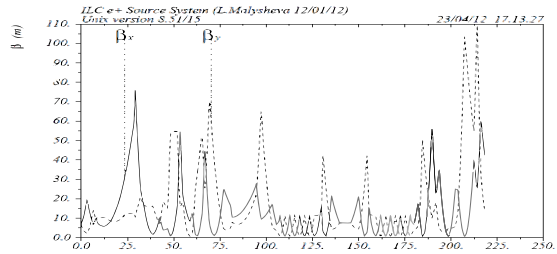}
   \caption{ Complete PLTR section including one of spin rotator branch  with solenoid 2.}
   \label{f4}
\end{figure}

The section  consists of the first irregular FODO-like cell with a pulsed kicker and a combined function defocusing/bending  magnet,  followed by 4 regular FODO cells with $120^0$  phase advance, forming  together an achromat dogleg, a  solenoid matching section and a solenoid with an integrated field of 26.18 or 26.2 $[T \cdot m ]$.  In the  solenoid  beta functions $\beta_{x}=\beta_{y}$ and  they are reaching  the minimum in the middle of the solenoid.   The rest of the section is a mirror image of the first part with respect to the middle of solenoid. The second branch of the lattice can be obtained by switching the  sign of the kick in the pulsed kicker and the bending angles in the following dogleg. The section was optimized by MAD8 package \cite{10} to meet the constraints on the length of total D section.
   Then this spin-rotator part of  section D was inserted to the PLTR lattice developed by W.Liu \cite{11} thus including  two extra matching sections. In Fig.~\ref{f3}-Fig.~\ref{f4} the results of the optics  matching using MAD8 package  is given  for PLRT section. The lattice  matching for the section D of the PLRT was done for both  versions of  superconducting solenoids.

 The  optics for Solenoid 1 setting  is  cross-checked  with ELEGANT  code\cite{12} code. Spin tracking  with BMAD code \cite{13} is done  by V.Kovalenko \cite{14}.

 \subsection{Asymmetric Design}

\begin{figure}[htb]
   \centering
   \includegraphics*[width=130mm,height=40mm]{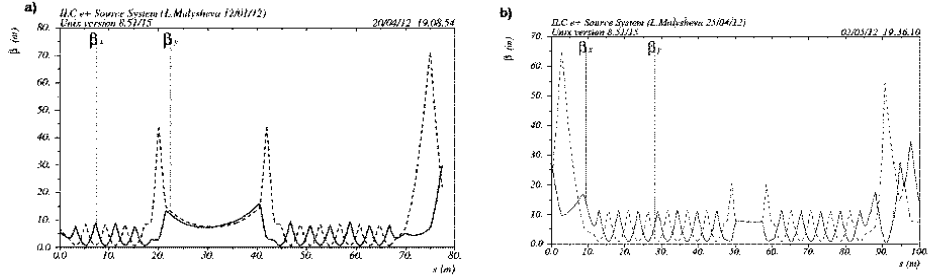}
   \caption{a) Asymmetric section (without  energy compressor) for the spin rotator branch with Solenoid 1. b)Asymmetric section with the energy compressor part for  the spin rotator branch with Solenoid 2 (b). }
   \label{f5}
\end{figure}

    Two solenoid sections in the opposite branches were placed with  $\approx 6-11$ m shift producing  a horizontal offset of 0.54 m for each branch with respect to design orbit.   One or two extra FODO cells were added  in front of  the solenoid section for one branch and  the same number of  extra FODO cells  were added after the solenoid section for another branch.   These changes  lead to  an increase of  the  total length of the whole spin rotator section.   Rematching  was done in order to fit the  total length of section D (123.595m) and the total PLTR length. In Fig.~\ref{f5} the design of the  new spin rotation section  for two version of super-conducting solenoids is given.

\section{CONCLUSIONS}
 The  suggested spin rotator design  confirms that   fast helicity switching for the positron beam is possible.   The  train-to-train polarity selection  for electron and positron beams at the IP  can be achieved.  In particular:
\begin{itemize}
    \item  The  suggested optic design  for the  fast helicity reversal spin rotator section satisfies  the PLTR section requirements.
    \item   Both  versions of superconducting solenoid for spin rotator are used and two versions of  optic files are available for the symmetric lattice
    \item  An asymmetric design for the solenoid position  in two parallel lines of spin rotator is produced for two version of superconducting solenoid parameters.
    \item  The  optic  design is cross-checked with  different accelerator design codes
    \item   Depolarization effects in a new lattice are estimated by BMAD  and no significant  depolarization connected with beam optics is discovered.
\end{itemize}

\section{ACKNOWLEDGMENT}
The authors are grateful to  the all members   of the Spin-management group  for the fruitful discussions. We  would like to thank Dr. W. Liu  for close collaboration and for providing the matching  parameters of the PLTR lattice.  L.I. Malysheva also thanks Mr. J. Jones and Dr. P. Williams from ASTeC for help and advice concerning  ELEGANT running.

\end{document}